\documentclass[12pt,preprint]{aastex}
\shortauthors{Ramanpreet Kaur et al.}
\shorttitle{QPO evolution in 4U 1626--67}

\begin{document}
\title{A study of the long term evolution of quasi periodic
oscillations in the accretion powered X-ray pulsar 4U 1626--67}

\author{Ramanpreet Kaur\altaffilmark{1},
 Biswajit Paul\altaffilmark{2}, Brijesh Kumar\altaffilmark{1,3}, Ram Sagar\altaffilmark{1}}

\altaffiltext{1}{Aryabhatta Research Institute of Observational Sciences, Manora Peak, NainiTal 263\,129, India}
\altaffiltext{2}{Raman Research Institute, C.V. Raman Avenue, Sadashivanagar, Bangalore 560\,080, India}
\altaffiltext{3}{Departamento de Fi$\!^{'}\!$sica, Universidad de Concepcio$\!^{'}\!$n, Casilla 160-C, Concepcio$\!^{'}\!$n, Chile}

%\maketitle

\label{firstpage}

\begin{abstract}

We report here a study of the long term properties of Quasi Periodic
Oscillations (QPO) in an unusual accreting X-ray pulsar, 4U 1626--67. This
is a unique accretion powered X-ray pulsar in which we have found the
QPOs to be present during all sufficiently long X-ray observations with
a wide range of X-ray observatories. In the present spin-down era of this
source, the QPO central frequency is found to be decreasing. In the earlier
spin-up era of this source, there are only two reports of QPO detections, 
in 1983 with EXOSAT and 1988 with GINGA with an increasing trend.
The QPO frequency evolution in 4U 1626--67 during the last 22 years changed
from a positive to a negative trend, somewhat coincident with the torque 
reversal in this source. In the accretion powered X-ray pulsars, the QPO frequency
is directly related to the inner radius of the accretion disk, as per 
Keplerian Frequency Model (KFM) and Beat Frequency Model (BFM). 
A gradual depletion of accretion disk 
is reported earlier from the X-ray spectral, flux and pulse profile measurements.
The present QPO frequency evolution study shows that X-ray flux and mass accretion 
rate may not change by the same factor, hence the simple KFM and BFM are not able to
explain the QPO evolution in this source.
This is the only X-ray pulsar to show persistent QPOs 
and is also the first accreting X-ray pulsar in which the QPO history is 
reported for a long time scale relating it with the long term evolution 
of the accretion disk.
 
\end{abstract}

\keywords{
binaries: close - pulsars: individual: 4U 1626--67 - stars: neutron - X-rays: binaries
}

\section{Introduction}

The X-ray source 4U 1626--67 was discovered with the Uhuru satellite
(Giacconi et al. 1972) in 2-6 keV band. 
Pulsations, with a period of 7.68 s were first discovered by
Rappaport et al. (1977) with SAS-3 observations and has been extensively
monitored since then, especially with the BATSE detectors onboard CGRO
(Chakrabarty et al 1997; Bildsten et al 1997). 
Optical counterpart of the pulsar was identified as KZ TrA, a faint blue star
(V $\approx$ 18.5) with little or no reddening (McClintock et al. 1977; Bradt and 
McClintock 1983). 
Optical pulsations with 2\% amplitude were detected at the same frequency
as the X-ray pulsations (Ilovaisky, Motch, \& Chevalier 1978) and are
understood to be due to reprocessing of the pulsed X-ray flux by the
accretion disk (Chester 1979). A faint optical counterpart and the observed optical 
pulsed fraction requires the companion star to be of very small mass
(McClintock et al. 1977, 1980). 
The X-ray light curve does not show any orbital modulation or eclipse.
However, from the reprocessed pulsed optical emission and a close sideband
in the power-spectrum of optical light curve, an orbital period of 42 minutes
was inferred (Middleditch et al. 1981). 
Therefore, it falls under the category of ultra compact binaries
(P$_{orb}$ $<$ 80 minutes), which have hydrogen-depleted secondaries to reach
such short periods (Paczynski \& Sienkiewicz 1981; Nelson et al. 1986).

Despite extensive searches, the orbital motion of this binary has never
been detected in the X-ray pulse timing studies (Rappaport et al, 1977;
Levine et al. 1988; Jain et al. 2007).
A very low mass secondary, in a nearly face on orbit can possibly account for
the lack of pulse arrival time delay. Recently Jain et al (2007)
have also proposed this source to be a candidate for a neutron star with
a supernova fall back accretion disk.
From the extensive timing and spectral observations both in optical and
X-ray bands, it has not yet been possible to establish the presence of a
binary companion, and the upper limit of the companion mass has been
determined to be very low.
However, the presence of an accretion disk in 4U 1626--67 is beyond any
doubt. Optical spectral and timing studies confirm that most of the optical 
emission is strongly dominated by the accretion disk (Grindlay 1978;
McClintock et al. 1980). The X-ray spectrum also shows bright hydrogen-like and
helium-like oxygen and neon emission lines with red and blue shifted
components, a certain sign for accretion disk origin (Schulz et al. 2001,
Krauss et al. 2007).
Another direct evidence of an accretion disk in 4U 1626--67 is found from the
detection of quasi-periodic oscillations, at a frequency of 40 mHz, from Ginga
observations (Shinoda et al. 1990) and subsequently at a higher frequency of
about 48 mHz from Beppo-SAX, ASCA, RXTE and XMM-Newton (Owens et al. 1997; 
Angelini et al. 1995; Kommers et al. 1998, Krauss et al. 2007). The QPOs have also
been detected in reprocessed optical emission from both ground based
and HST observations (Chakrabarty et al. 1998, 2001).

For more than a decade since its discovery, 4U 1626--67 was found to be
spinning up with a characteristic timescale P/\.{P}  $\approx$ 5000 yr.
It was found to be spinning down at about the same rate by BATSE onboard CGRO in
the beginning of 1991 (Chakrabarty et al. 1997). Even though 
the torque reversal was abrupt, the decrease in bolometric X-ray flux has been
gradual and continuous over the past $\approx$ 30 yr (Chakrabarty et al.
1997, Krauss et al 2007). Recently, from a set of Chandra monitoring
observations Krauss et al (2007) have established that the bolometric 
X-ray flux and various emission line fluxes
have decreased continuously over the last few years, indicating a gradual
depletion of the accretion disk. The X-ray flux and mass accretion rate are
directly related and these are likely to be related to the mass and extent
of the material in the accretion disk. Therefore, the observed gradual
decrease in X-ray flux indicates a depletion of material in the accretion 
disk of the pulsar. Another signature of this is seen by Krauss et al. (2007)
as a change in the pulse profile of the pulsar as compared to the
earlier observations. 
 
In the present work, we have investigated the QPO frequency evolution of
4U 1626--67 over a long period and discuss the relation of the change in
QPO frequency with the a possible recession of the inner accretion disk.

\section{Observations and Analysis}

4U 1626--67 has been observed with various X-ray telescopes over different
epochs of time. Table 1 lists the log of observations of 4U 1626--67 that
were found to be useful for the present study.
Details of individual observations described below are in chronological order.
Detection of QPOs at around 48 mHz have been mentioned from some of these
observations, sometimes from a different instrument also (Ginga - Shinoda
et al. 1990; ASCA - Angelini et al. 1995; Beppo-SAX - Owens et al. 1997,
RXTE - Kommers et al. 1998, Chakrabarty 1998; XMM-Newton - Krauss et al.
2007). However, the QPO frequencies measured from these observations are
often not reported with good enough accuracy to investigate a slow frequency
evolution. For the present study, we have therefore reanalysed the data
and measured the QPO parameters with the highest possible accuracy.\\

EXOSAT Medium Energy (ME) proportional counter lightcurve of 4U 1626--67
was obtained from HEASARC archive with the time resolution of 0.3125 s for
an observations made on August 30, 1983 for 27 ks. ME lightcurve of another
observations made by EXOSAT on March 30, 1986 for $\approx$ 84 ks that was 
reported earlier by Levine et al. (1988) is not available in the HEASARC 
Archive. \\

ASCA observations of 4U 1626--67 were made on August 11, 1993 with the
two Gas Imaging Spectrometers (GIS2 and GIS3) and the two Solid-state Imaging
Spectrometers (SIS0 and SIS1) and light curves with total useful exposures
of 40 ks and 25 ks were obtained for the GIS and SIS respectively. 
During the ASCA observation, the GIS detectors were operated in Pulse Height
mode and SIS detectors were operated in Fast mode and the lightcurves 
were extracted from the unscreened high bit mode data with the minimum time 
resolution of 0.125 s for both GIS and SIS detectors. The light curves from 
the pairs of GIS and SIS instruments were added and a single power spectra 
is generated with the summed lightcurves.\\ 

4U 1626--67 was observed with BeppoSAX on August 09, 1996 for 116 ks by the
three units of Medium Energy Concentrator Spectrometer (MECS) and for 35 ks
by the Low Energy Concentrator Spectrometer (LECS). Lightcurves were extracted
from all the instruments with 0.125 s. Single summed lightcurve was generated 
from three lightcurves of the MECS instruments to increase the signal-to-noise ratio.\\ 

RXTE-PCA pointed observations of the source were made from February 1996 to 
August 1998. In 1996, the observations were made in the beginning of the 
year and at the end of the year under obs ID P10101 and P10144 respectively. 
The observations made under obs IDs P10101 covers time span of almost 
5 days from MJD 50123 to 50128. There were nine observations in this 
obs ID each lasting for 4-8 hrs. A single observation was made under 
obs ID P10144 for $\approx$ 5 hrs on MJD 50445. In 1997, all the 
observations were made under obs ID P20146 covers a time range 
of almost a year from MJD 50412 to MJD 50795 but individual observations 
were made only for a few minutes. In 1998, RXTE-PCA made
observations under two obs ID P30058 and P30060. There were three
observations made under obs ID P30058, out of which two observations were
made on MJD 50926 and the third observation was made on MJD 51032. In
obs ID P30060, there were 10 short observations each for about an
hour. For almost all the observations of RXTE, all five PCUs were on.
Lightcurves were extracted from observations of 4U 1626--67 with a time
resolution of 0.125 s using the Standard-1 data that covers the entire
2-60 keV energy range of the PCA detectors. We divided the whole
RXTE-PCA observations from 1996 to 1998 into three segments from MJD 50123
to 50128, 50412 to 50795 and 50926 to 51032. The signal-to-noise ratio of
the power spectra generated from the individual observations made between
MJD 50412 to 50795 was poor to detect QPO except on MJD 50445, thus a single
power spectrum was produced by combining powerspectra of all observations 
made between MJD 50412 to 50795.\\

XMM-Newton has observed 4U 1626--67 four times, but significant amount of
science data was present only in two of these observations, made under obs IDs 
0111070201 and 0152620101, listed in Table 1.
We have analysed data only from PN detector of European Photon Imaging Camera
(EPIC) onboard XMM-Newton. PN operates in the energy band of 0.15-15 keV.
Lightcurves were extracted with a time resolution of 0.125 s for both the
observations. \\

All the lightcurves were divided into small segments each of length 1024 s
and a power density spectrum of each segment was generated.
The power spectra were normalized such that their integral gives the standard
rms fractional variability and the expected white noise was subtracted.
Final power spectra was generated with the average of all the power spectra 
generated for each of the observations listed in Table 1.
Flares with duration of 1000 s are clearly seen in the EXOSAT data as mentioned by
Levine et al. (1988). However these flares are not detected in rest of the data 
mentioned in Table 1.
QPO at a frequency of $\sim$ 48 mHz is clearly seen in the power spectra
of all the data sets except from EXOSAT observations during which it is detected at $\sim$
36 mHz. Figure 1 shows the QPO detection from the EXOSAT observations made on August 30, 1983
in the range of 15 mHz to 100 mHz. A Gaussian model is fitted to the QPO feature to
determine its central frequency and width (FWHM of Gaussian) for all the datasets.
The continuum of the power spectrum in the band of 20 mHz to 80 mHz is fitted with a constant
or a linear model. The uncertainty of the Gaussian model peak at 1 $\sigma$ confidence 
interval is quoted as an error on the Gaussian centre. 

The QPO feature detected in the power spectrum of EXOSAT data is quite narrow $\sim$
2 mHz as compared to the QPOs seen in rest of the data with a width of $\sim$ 4 to 5 mHz.
Figure 2 shows powerspectra in the frequency range 26 mHz to 72 mHz for the observations
listed in Table 1 except the EXOSAT observations. Different constant numbers were added 
to each plot for clarity. A best-fitted Gaussian model for the QPOs 
and a constant model or a linear model for the continuum is shown on each plot with 
a solid line. A dotted vertical line at the best 
fitted Gaussian center to the ASCA 1993 data is plotted in the same figure. A shift of $\sim$
2 mHz is clearly seen from bottom to the top plot shown in Figure 2. 

The evolution of the QPO central frequency as observed by 
various X-ray telescopes in both spin-up and spin down era is shown in Figure 3. 
An error bar plotted on each point in Fig 3 represents 1$\sigma$ error estimates.
We couldn't find GINGA observations of 4U 1626--67 made in July, 1988 from archive data, thus the
central frequency of QPOs and error estimate on it is taken from Shinoda et al. 1990
and is also shown in Fig 3. To confirm the consistency of QPO frequency for each 
data set listed in Table 1, the QPO frequencies were measured from smaller 
segments of the data, 10 each for the 1996 RXTE observation and the 
2004 XMM observation. The values determined from smaller segments 
have larger uncertainties but within uncertainties, these values are consistent
with the QPO frequency measured using the complete data sets in each case.
It can be clearly seen in Figure 3 that the QPO central frequency has increased from
1983 to 1993 and after that it gradually decreased from 1993 to 2004. However the lack 
of observations doesn't allow us to define an exact time when the QPO frequency 
evolution changed from an increasing trend to a decreasing trend. The  
observations from 1993 to 2004 showed frequency decrease of $\sim$ 2.3 mHz while 
the error bars on all the data points during this era are within 0.4 mHz except the 
ASCA 1993 data point for which the error bar is 0.6 mHz, confirms the real decrease 
in QPO frequency with time. The QPO frequency derivative during spin-down era is 
$\sim$ (0.2 $\pm$ 0.05) mHz/yr. A linear fit is  shown on the data points with a solid line 
in the spin-down era in Figure 3. The reduced $\chi^2$ of the linear 
fit is 1.07 for 5 degrees of freedom. To further confirm the linearity, a constant model 
is also fitted to the data from 1993 to 2004. The reduced $\chi^2$ for a constant model is 3.22 
for 6 degrees of freedom, indicates poor fit as compared to the linear fit. 

\section{Discussion}

In high magnetic field X-ray pulsars, the QPO frequency is in the range of
a few mHz to a few Hz (Kaur et al. 2007). The QPOs are known to occur
sporadically, only in a few percent of the X-ray observations.
For example, QPOs are detected in only 15\% of the out-of-eclipse
observations of Cen X-3 (Raichur et al. 2007). Our independent investigation
of the RXTE-PCA lightcurves of several persistent sources show that the
QPOs are quite rare. Exception to this are some of the transient sources,
like 3A 0535+262 (Finger et al. 1996), and XTE J1858+034 (Paul \&
Rao 1998) which showed QPOs during most of the observations made during
their outbursts. In the present study, using lightcurves of 4U 1626--67
taken with various observatories over a period of more than 20 years we
have detected QPOs in every single observation of sufficient length.
This is the first accretion
powered pulsar for which the QPO study has been made over a long time scale.
In this regard, 4U 1626--67 is unique among persistent high magnetic field
accreting X-ray pulsars. It shows that the accretion disk of the pulsar is
quite stable to hold this feature for years. However, in a few cases, the
observation duration was not long enough to make accurate measurement of
the QPO parameters.  

QPOs in accretion powered  X-ray sources are widely believed to arise
due to inhomogeneities near the inner accretion disk. The QPO frequency
is the Keplerian frequency at the inner disk radius and is therefore positively
related to the mass accretion rate or the X-ray luminosity. If the compact
object is a neutron star, the inner disk is coupled with the central object
through the magnetic field lines and QPOs corresponding to the beat
frequency between the spin frequency and the Keplerian frequency of the
inner disk can also be seen. In accretion powered high magnetic field X-ray pulsars,
the two different QPOs are never seen to occur in the same source.
In some of the sources, like 4U 1626--67, the QPO frequency is lower than
the spin frequency and therefore the QPOs can only be explained by the
BFM.

According to both KFM and BFM, the radius of the QPO production 
area, r$_{qpo}$, is defined as
\begin{equation}
r_{qpo} = \left(\frac{GM_{NS}}{4\pi^2\nu_k^2}\right)^{1/3} 
\end{equation}
where G is the Gravitational constant, M$_{NS}$ is the mass of the neutron star and
$\nu_k$ is the keplerian frequency of the inner accretion disk. 

The radius of the inner accretion disk, r$_M$ can be defined as
\begin{equation}
r_M = 3 \times 10^8 L_{37}^{-2/7}\mu_{30}^{4/7}
\end{equation}
where L$_{37}$ is the X-ray luminosity in units of 10$^{37}$ ergs s$^{-1}$ and $\mu_{30}$ is 
magnetic moment in units of 10$^{30}$ cm$^3$Gauss. 
If the QPOs are as per Keplerian frequency model ($\nu_k$ = $\nu_{qpo}$, where $\nu_{qpo}$ is
QPO frequency of the pulsar), then we expect 
$\nu_k$ $\propto$ $L_{37}^{3/7}$ or $\nu_{qpo}$ $\propto$ $L_{37}^{3/7}$. The flux of 
4U 1626--67 has decreased from 0.32 to 0.15 units from 1993 to 2004 (Krauss et al. 2007), 
implies that the change in QPO frequency is expected to be $\sim$ 27\% from 1993 to 2004. 
The present QPO observations have shown only 4 \% decrease in QPO frequency during the 
same time. However, Keplerian frequency model is not valid in this source. In the BFM 
($\nu_k$ = $\nu_{qpo}$ + $\nu_s$, where $\nu_s$ is pulsar spin frequency), the inner disk 
frequency is higher as compared to 
KFM, and the relative change in QPO frequency is expected to be even larger. Therefore, 
we see that the evolution of QPO frequency and the decrease of X-ray flux cannot be 
explained in the standard QPO generation mechanism and usual relation between inner 
disk and X-ray luminosity.  We can consider two possibilities : One is that the QPOs 
are not generated from the inner disk, these are generated due to reprocessing in 
some outer structure of the disk. This is not very likely due to the large (upto 
15\%) rms in the QPO feature. Second possibility is that the observed X-ray flux 
change is not due to change of mass accretion rate by the same factor. Many X-ray 
sources show X-ray flux variation at long time scale upto a few months due to obstruction 
provided by complex accretion disk mechanism.

The earlier study by Chakrabarty et al. (1997)  has concluded that 
there was an abrupt torque reversal in 1990 and the system moved from 
spin-up to spin-down era with a characteristic time scale P/\.{P}  of 
$\sim$ 5000 yr. The two QPO detections with EXOSAT (35 mHz in 1983) 
and GINGA (40 mHz in 1988) are during the spin-up era of this pulsar, with increasing 
trend while the observations from 1993 to 2004, in the spin-down era, showed 
a slow decreasing trend in QPO frequency with time, somewhat coincident 
with the torque reversal in this source, shown in Fig 3. QPO 
frequency is found to be decreasing in the spin-down era with a frequency derivative 
of $\sim$ (0.2 $\pm$ 0.05) mHz/yr. 
The X-ray spectral and flux evolution study along with pulse profile changes
of 4U 1626--67 by Krauss et al (2007) have concluded that the accretion disk
in this source is depleting with a time scale of 30-70 years. Krauss et al.
(2007) has also estimated the long term average accretion rate to be
3 $\times$ 10$^{-11}$ M$_\odot$ yr$^{-1}$ for a distance $\ge$ 3kpc.
However, a gradual change in mass accretion rate 
can not explain the unique torque reversal phenomena of this source (Li et al. 1980).

\section{Conclusions}

\begin{itemize}
\item We have detected very persistent quasi-periodic oscillations in the unique
accretion powered X-ray pulsar 4U 1626--67.
\item Using data from several observatories, we have detected a gradual
evolution of the oscillation frequency over a period of 22 years.
\item The frequency evolution indicates a possible recession of the accretion disk
of the pulsar during the present spin-down era.
\end{itemize}

\section*{Acknowledgments} 
This research has made use of data obtained through the High Energy Astrophysics 
Science Archive Research Center Online Service, provided by 
the NASA/Goddard Space Flight Center.

\newpage
%Tables

\clearpage
{
\begin{table}
\caption{Log of Observations of 4U 1626--67}
\begin{tabular}{|c|c|c|c|c|c|}
\hline
Telescope&Year&Obs Ids&No. of & Obs span (ks)         & Time on \\
         &    &       & Pointings               &(End time--Start time) & Source (ks)  \\
\hline
EXOSAT/ME   &1983&128       &1&27&27\\
\hline
ASCA/GIS    &1993&40021000&1&72&40\\
ASCA/SIS    &1993&40021000&1&70&25\\
\hline
BeppoSAX/MECS &1996&10017001&1&162&116\\
BeppoSAX/LECS &1996&10017001&1&128&35\\
\hline
RXTE/PCA &1996&P10101&9&395&147\\
\hline
RXTE/PCA &1996&P10144&1&13&10\\
         & 1997&P20146&14&33125&13\\
\hline
RXTE/PCA      &1998&P30058&3&9167&40\\
         &    &P30060&10&2758&44\\
\hline
XMM-Newton/PN  &2001&0111070201&1&17&16\\
\hline
XMM-Newton/PN  &2003&0152620101&1&84&84\\
\hline
\end{tabular}
\end{table}
}
\clearpage
\begin{figure}
\centering
\includegraphics[height=6.5in, angle =-90]{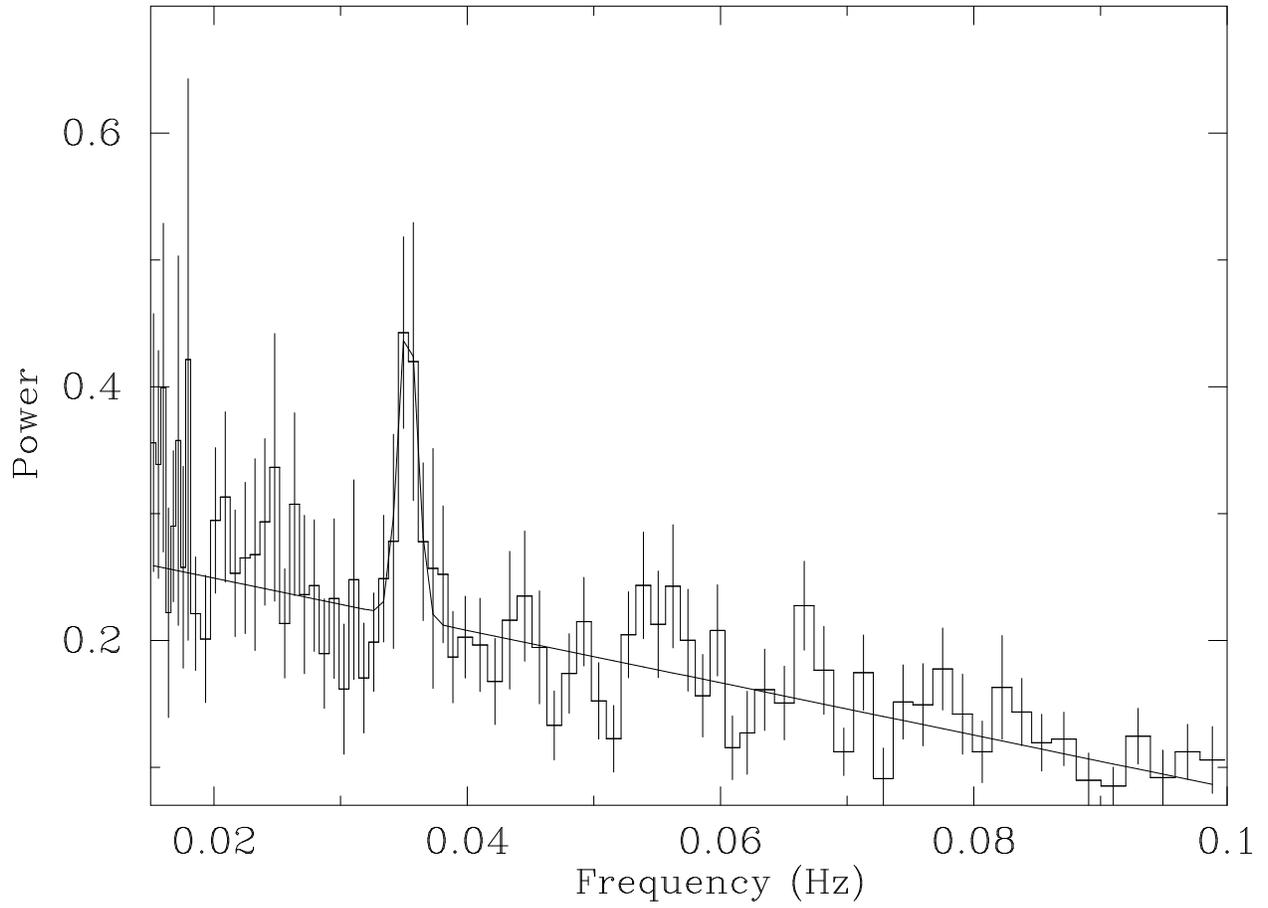}
\caption{Power density spectrum generated from the lightcurve obtained from
the EXOSAT observation made on August 30, 1983.}
\end{figure}

\clearpage

\begin{figure}
\centering
\includegraphics[height=5.5in, angle =-90]{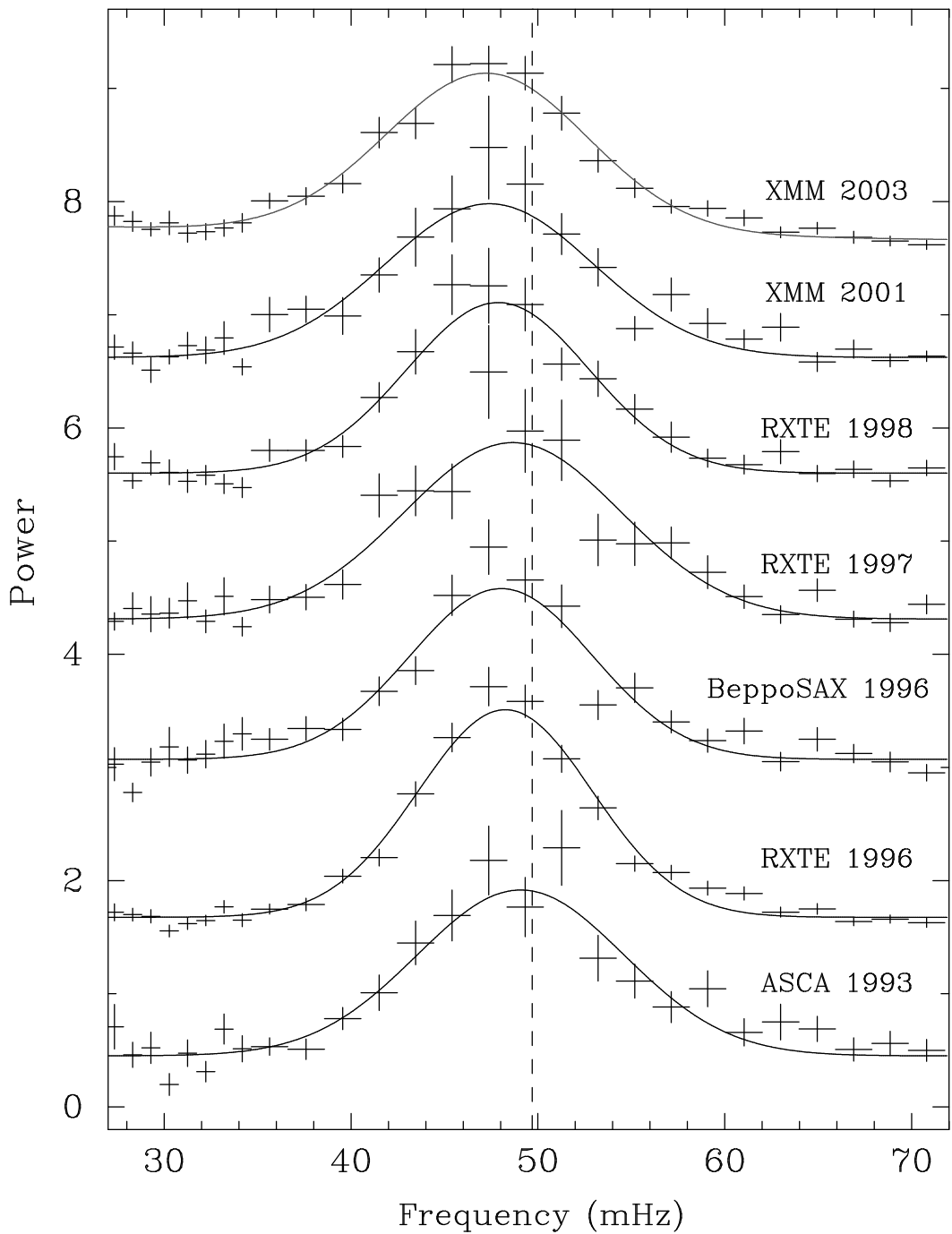}
\caption{All power density spectrum are generated from the lightcurves obtained 
from observations listed in Table 1. in chronological order. Different constant 
numbers were added to each plot for clarity. The year of observations is written along with 
each PDS. A vertical line is drawn at 49.77 mHz, QPO frequency of ASCA 1993 observations,  
to clearly see the decrease in QPO frequency with time.}
\end{figure}

\clearpage

\begin{figure}
\centering
\includegraphics[height=6.5in, angle =-90]{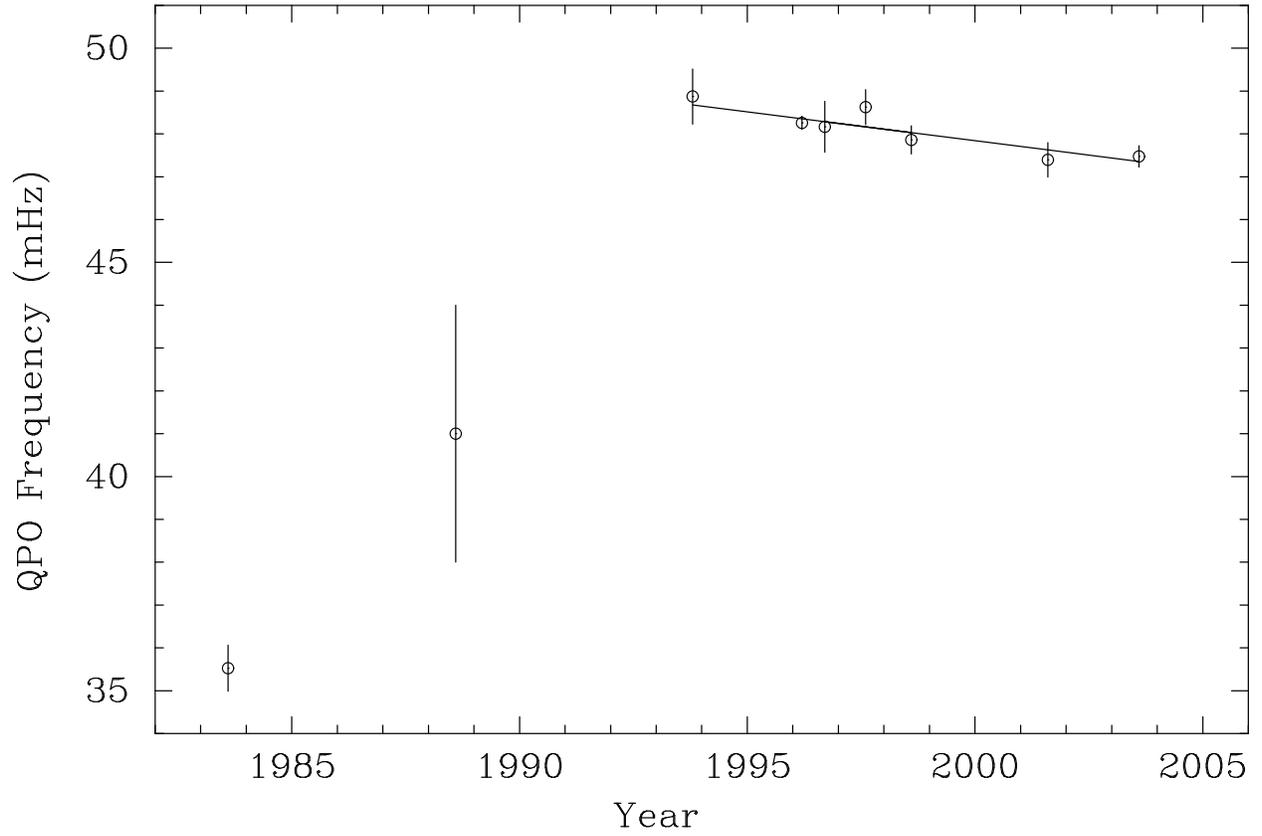}
\caption{QPO frequency evolution history of 4U 1626--67 from 1983 to 2004.
The solid line is a linear fit to the data from 1993 to 2004. Error bars represent the
1$\sigma$ confidence intervals.}
\end{figure}

\end{document}